\documentclass{eas}
\usepackage{graphicx}
\begin{document}

\title{Tracing FUV Radiation in the Embedded Phase of Star Formation}
\runningtitle{Tracing FUV Radiation}
\author{Arnold O. Benz}\address{Institute of Astronomy, ETH Zurich, 8093 Z\"urich, Switzerland}
\author{Simon Bruderer}\address{Max Planck Institut f\"{u}r extraterrestrische Physik, Giessenbachstrasse 1, 85748 Garching, Germany}
\author{Ewine F. van Dishoeck$^{2,}$}\address{Leiden Observatory, Leiden University, PO Box 9513, 2300 RA Leiden, The Netherlands, and Max Planck Institut f\"{u}r extraterrestrische Physik, Giessenbachstrasse 1, 85748 Garching, Germany}
\author{Pascal St\"auber$^1$}
\author{Susanne F. Wampfler$^1$}
\author{Carolin Dedes$^1$}
\begin{abstract}
 Molecules containing one or a few hydrogen atoms and a heavier atom (hydrides) have been predicted to trace FUV radiation. In some chemical models, FUV emission by the central object or protostar of a star forming region greatly enhances some of the hydride abundances. Two massive regions, W3 IRS5 and AFGL 2591, have been observed in hydride lines by HIFI onboard the {\it Herschel Space Observatory}. We use published results as well as new observations of CH$^+$ towards W3 IRS5. Molecular column densities are derived from ground state absorption lines, radiative transfer modeling or rotational diagrams. Models assuming no internal FUV are compared with two-dimensional models including FUV irradiation of outflow walls. We confirm that the effect of FUV is clearly noticeable and greatly improves the fit. The most sensitive molecules to FUV irradiation are CH$^+$ and OH$^+$, enhanced in abundance by many orders of magnitude. Modeling in addition also full line radiative transfer, Bruderer et al (2010b) achieve good agreement of a two-dimensional FUV model with observations of CH$^+$ in AFGL 2591. It is concluded that CH$^+$ and OH$^+$ are good FUV tracers in star-forming regions.
\end{abstract}
\maketitle
\section{Introduction}
The formation process of high-mass stars is still disputed. In the scenario of continuous accretion, the young stellar object (YSO) starts to burn deuterium before reaching the main sequence. When this burning reaches the outer shells, the YSO inflates and the surface temperature is reduced (Behrend \& Maeder 2001). The same happens if the accretion rate exceeds $10^{-3} M_\odot$ yr$^{-1}$ and YSO radii may exceed 100 $R_\odot$ (Hosokawa, Yorke, \& Omukai 2010). A continuously accreting YSO moves in the Hertzsprung-Russel diagram above and parallel to the main sequence. As the temperature reduction depends on the accretion rate, it depends on initial circumstances when the surface exceeds $10^4$ K, the temperature at which a significant fraction of the YSO luminosity is emitted in ionizing UV radiation. Alternatively, such radiation can be produced by accretion or jet driven shocks. X-ray emission of the YSO is limited to a small fraction of the {\it Herschel} beam for the distant massive objects studied here and is overwhelmed by far-UV (FUV, 6 - 13.6 eV) if present (Bruderer et al. 2010a).

We present results of the subprogram `Radiation Diagnostics' of the {\it Herschel} Key Program `Water in Star-forming regions with {\it Herschel}' (WISH, van Dishoeck et al 2011). The goal of the subprogram is to explore the possibilities of identifying FUV and X-ray emission through chemistry in deeply embedded objects, where UV and X-rays cannot be observed directly due to a high attenuating column density. Key questions are to quantify such high-energy radiation and their influence on star- and planet-forming regions.

Here we test molecular UV tracers in massive YSOs known to have Ultra Compact H II regions indicating that their UV emission has begun. We focus on hydrides, such as OH, CH, NH, SH, H$_2$O and their ions, OH$^+$, CH$^+$, NH$^+$, SH$^+$, H$_2$O$^+$, and H$_3$O$^+$. Many hydrides have a high activation energy in their formation reactions. However, high-energy photons such as FUV or X-rays heat and ionize the molecular gas. The hydride abundances are greatly enhanced by the high temperatures (Sternberg \& Dalgarno 1995). Ionized hydrides are chemically active and can drive substantial chemical evolution. If their chemistry and excitation is understood better, they may become valuable tracers of warm and ionized gas and of the embedded phase in star and planet formation.

\begin{table}[htb]
\begin{center}
\resizebox{6.5cm}{!}{
\begin{tabular}{lrcc}   
\hline \hline
Mole- &Model& Model&Ratio \\
cule &FUV& no-FUV &FUV/no-FUV \\
&in units of $10^{-12}$&in units of $10^{-12}$& \\
\hline\\
CH&2000&20&100 \\
NH&900&60&15\\
SH&20&3000&0.006 \\
OH$^+$&3000&0.004&7$\times 10^5$ \\
CH$^+$&4000&6$\times 10^{-5}$&7$\times 10^7$ \\
NH$^+$&30&3$\times 10^{-4}$&1$\times 10^5$  \\
SH$^+$&200&2&100  \\
H$_2$O$^+$&500&0.01&$5\times 10^4$ \\
H$_3$O$^+$&900&600&1.5 \\
\hline
\end{tabular}}
\end{center}
\caption{Volume averaged (8000 AU diameter) fractional abundances relative to hydrogen (H$_{\rm tot}$) derived from chemical model calculations. The no-FUV model is spherically symmetric, includes cosmic rays, but no internal FUV irradiation. The FUV model is two-dimensional and assumes protostellar FUV irradiating the envelope including the walls of the outflows (from Bruderer et al. 2010a).}
\label{table1}
\end{table}

\section{Model}
Molecular abundances were calculated in a time-dependent chemical network irradiated by FUV and X-ray radiation (Hollenbach \& Tielens 1999; St\"auber et al. 2007; Bruderer et al. 2009). Here we use the template density models of AFGL 2591 (van der Tak et al. 1999) and calculate the dust temperature from irradiation. We assume a chemical age of 4$\times 10^4$ yr and a YSO FUV luminosity of 4$\times 10^{37}$ erg s$^{-1}$ corresponding to a bolometric luminosity of 2$\times 10^4\ L_\odot$ and to an effective temperature of $3\times 10^4$ K. Abundances are determined in time and radial position. Finally the abundance is averaged over a volume weighted by density (Tab. \ref{table1}).

Two methods to test theory with observations are used here. In a first and simple approach, the molecular abundances are estimated from observations and compared with models. In a second approach, the line emission of the molecules at the derived abundance are modeled by radiation transfer, convolved with the {\it Herschel} beam, and directly compared with observed line strengths.

Two models are used: In the `no-FUV' model, only cosmic rays ionize the infalling envelope without additional high-energy irradiation by the YSO. It is a one-dimensional model with radial dependence on temperature and density, where the temperature model of Doty et al. (1999) is used. It assumes a cavity of 200 AU radius in the interior. The `FUV' model includes FUV emission by the central object(s) and assumes irradiation of the walls carved out by the outflows. The modeled outflow has negligible column depth and the shape of the cavity allows direct irradiation of the walls. This irradiation both ionizes and heats a thin layer out to a distance beyond the {\it Herschel} beam on scales of 10000 AU.

\begin{figure}[htb]
\centering
\resizebox{6cm}{!}{\includegraphics[angle=270]{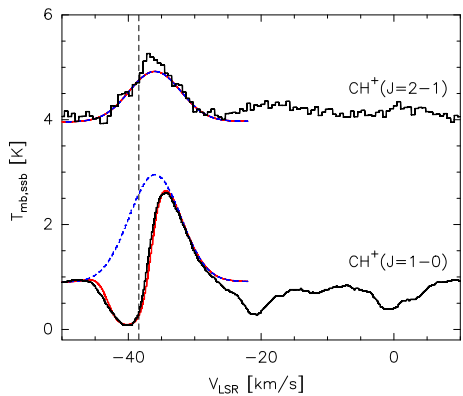}}
\caption{Line transitions of CH$^+$ observed with {\it Herschel/HIFI} towards W3 IRS5. Zero LSR velocity corresponds to the LSR frequency of 835.1375 GHz (M\"uller 2010).  The systemic velocity of W3 IRS5, -38.4 km s$^{-1}$, is shown by a vertical dashed line. The blue dashed curve indicates the contribution of the emission component. The red curve combines the contribution of the emission and absorption components fitted by the slab model.}
\label{CH+1-0}
\end{figure}

\section{Observations}

Hydride observations of two massive YSOs, W3 IRS5 and AFGL 2591, were reported by Benz et al. (2010) and Bruderer et al. (2010b), respectively. Figure \ref{CH+1-0} presents an additional observation for the first time. In both sources, the lines of CH$^+$ have a P Cyg profile in the $J=1-0$ transition and an emission peak in the $J=2-1$ transition at the systemic velocity. Combining the two lines, the abundance and temperature of the emission component, and the abundance of the ground state level of the absorbing component can be estimated using a slab model (Bruderer et al. 2010b). It is assumed that the absorption region is in front of the emission region, possibly associated with an outflow of the YSO. The best fitting model, indicated in Fig. \ref{CH+1-0}, yields a column density of $9.6\times10^{12}$ cm$^{-2}$ at an excitation temperature of 43 K for the emission component, and $4.4\times 10^{13}$ cm$^{-2}$ for the absorption component referring only to the ground state. Note that in both W3 IRS5 and AFGL 2591 the CH$^+$ $J=2-1$ line is relatively narrow ($\Delta V \approx 7 $km s$^{-1}$) and symmetric unlike emission expected from shocked regions. Emissions and absorptions at $V_{LSR} > -25$ km s$^{-1}$ may be caused by foreground clouds and are not modeled.

\section{Comparison of Observations and Theory}

\subsection{Comparison of abundances}
Column densities $N(x)$ are extracted by integrating line fluxes of a species $x$, neglecting re-emission or re-absorption of the final state. For H$_3$O$^+$  the total column density summed over all levels using the rotational diagram is derived (Benz et al. 2010). Only the line component corresponding to the systemic motion of the YSO and the emission component of CH$^+$ are used in the comparison.

\begin{figure}[htb]
\centering
\resizebox{12cm}{!}{\includegraphics{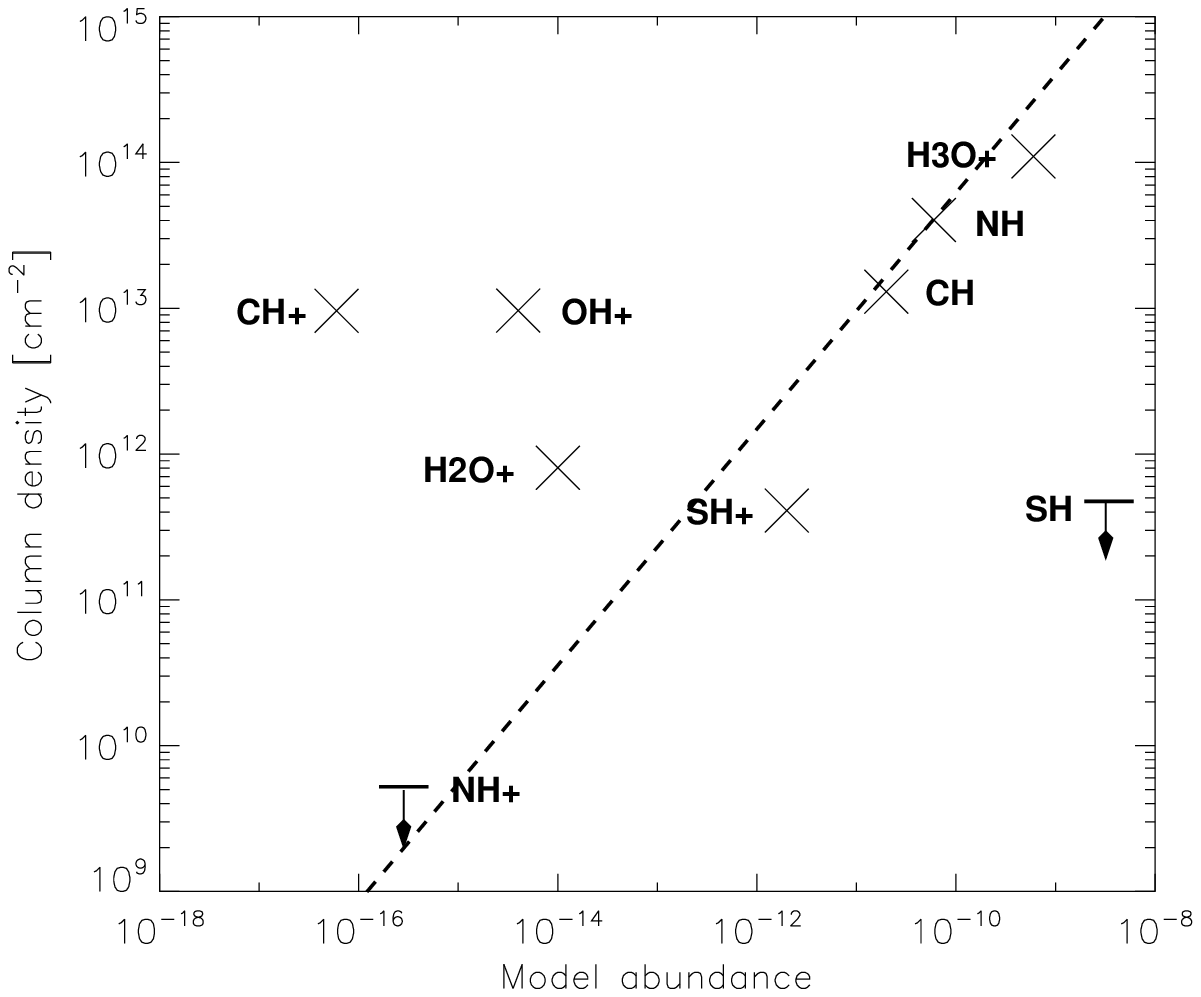}, \includegraphics{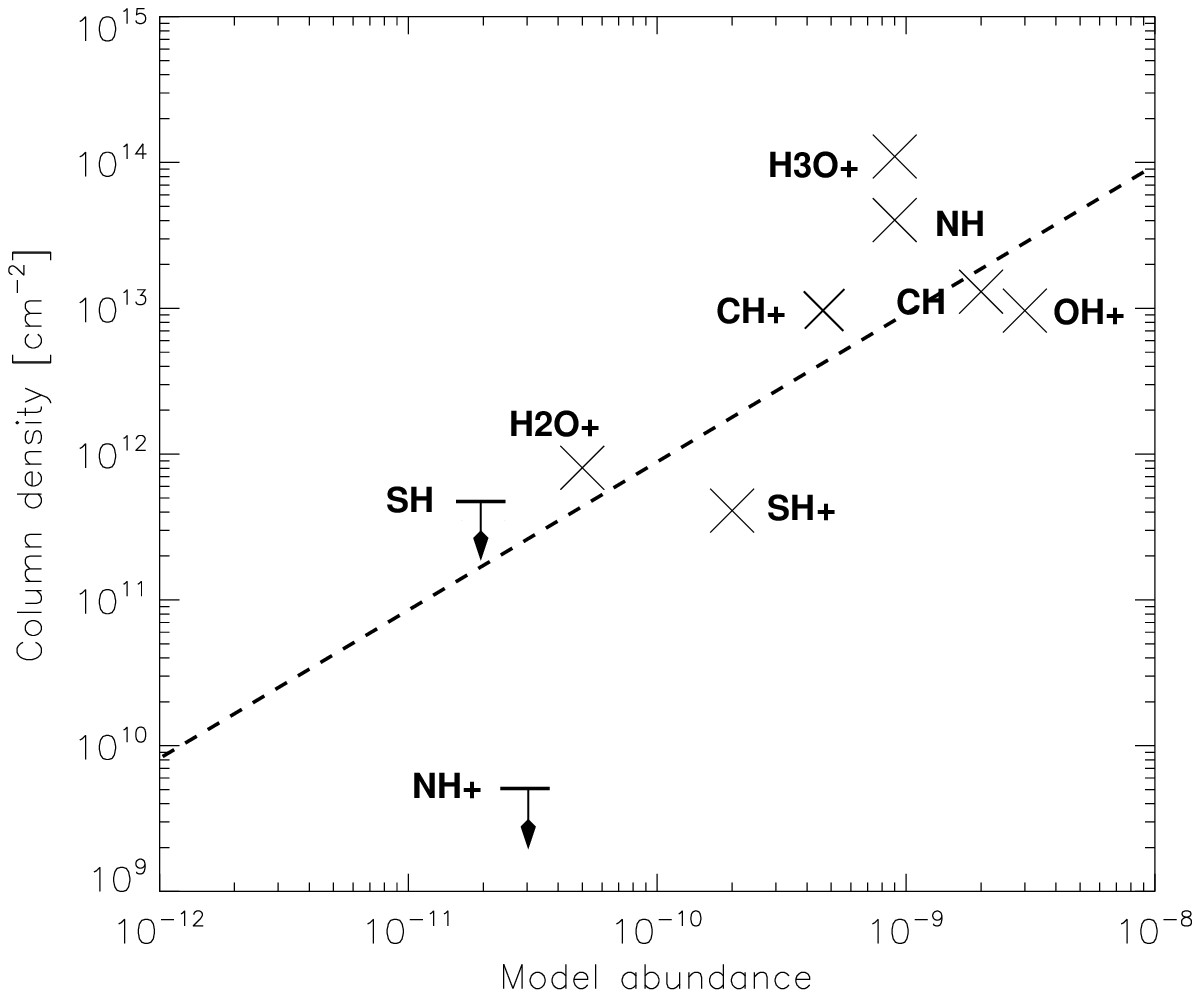}}
\caption{Comparison of column density derived from observations with model abundances calculated by Bruderer et al. (2010a) from a chemical model. A linear fit (see text) is indicated by the dashed line. {\it Left:} without internal FUV irradiation; {\it right:} internal FUV irradiation with a luminosity of 4$\times 10^{37}$ erg s$^{-1}$ . }
\label{comparison}
\end{figure}

In Fig. \ref{comparison} observations are compared to model predictions of volume averaged abundances $\chi(x)$ (Bruderer et al. 2010a). A dashed line  $N(x)$ = a $\chi(x)$ (where $a$ is considered a free parameter) is drawn such that the number of points above and below the line are equal (not counting upper limits above). In Fig. \ref{comparison} (left), the  column densities derived from observations are compared with the no-FUV model. If FUV is included in a spherical model, some molecules, such as CH$^+$ and OH$^+$ are enhanced an order of magnitude in abundance (St\"auber et al. 2005). However, the average  abundances of these molecules increase another five orders of magnitude or more if the FUV irradiates additionally the walls of the outflow cavity (Tab. \ref{CH+1-0}). The 2D model including FUV irradiation of the outflow walls is shown in Fig. \ref{comparison} (right). It fits significantly better.


\subsection{Comparison of line intensities}

\begin{table}[htb]
\begin{center}
\resizebox{10cm}{!}{
\begin{tabular}{lrrr}   
\hline \hline
&Abundance&Line flux (1-0)&Line flux (2-1)  \\
&&at 835.1375 GHz&at 1669.2813 GHz\\
&relative to H$_{\rm tot}$&[K km s$^{-1}$] &[K km s$^{-1}$]\\
\hline\\
Model no-FUV&$5\times 10^{-16}$&$\ll 0.001$ &$\ll 0.001$\\
Model FUV& $0.4 - 4\times 10^{-11}$&4.3 - 16.1&8.9 - 30.8\\
Observed& $9\times 10^{-11}$&0.91$\pm$0.03&3.7$\pm$0.2 \\
\hline
\end{tabular}}
\end{center}
\caption{Comparison of AFGL 2591 chemical and radiation transfer model calculations with {\it Herschel/HIFI} observations. The volume averaged model abundance refers to a radius of 20000 AU. The range of the FUV model indicates the range of free parameters  (from Bruderer et al. 2010b)}
\label{table}
\end{table}

Bruderer et al. (2010b) have modeled the CH$^+$ line emission of AFGL 2521 in the respective {\it Herschel} beam at 835 GHz and 1669 GHz (26.3'' and 12.7''). FUV irradiation in a two-dimensional geometry enhances the strengths of the CH$^+$ $J=1-0$ and $J=2-1$ lines close to the observed values. The agreement is well within the range given by the uncertainties of the chemical reaction rates and of the model assumptions (Table \ref{table}).

\section{Discussion and Conclusions}
Hydride observations of FUV emitting YSOs by HIFI onboard {\it Herschel} fit models including FUV irradiation considerably better than models without. Hence, FUV irradiation can be identified from molecular tracers. In star forming regions, CH$^+$ and OH$^+$ are the most FUV-sensitive species, and thus the best tracers. Therefore, the limitation to a single FUV tracer, such as CH$^+$ or OH$^+$, may be sufficient. Here we have emphasized CH$^+$ which is more enhanced and easier to detect. CH$^+$ emission probes the dense envelope; the order of magnitude fit of model and observations strongly suggest that CH$^+$ line emission is a method to trace enhanced FUV in star-forming regions.

OH$^+$ was observed mostly in absorption in W3 IRS5 and AFGL 2591, and only W3 IRS5 has a small P Cygni type emission. In Fig. \ref{comparison} the OH$^+$ absorption component was used. A proper excitation analysis of that molecule would be needed for use as a reliable tracer. OH$^+$ refers to more diffuse and lower density gas.

The other method compares abundances. Omitting effects of radiation transfer is clearly a coarse simplification. However, it includes contributions of various species that may originate from different locations and be better suited for quantifying the FUV flux in more complicated situations.

We emphasize that the effect of FUV irradiation is strongly enhanced by a two-dimensional geometry increasing the irradiated surface area (Bruderer et al. 2009). In these regions, the huge impact of FUV on the hydride chemistry in high-mass star forming regions is is not only caused by ionization, but also the result of gas heating.

\begin{acknowledgements}
We thank the WISH team for inspiring discussions and support. HIFI has been designed and built by a consortium of institutes and university departments from across Europe, Canada and the United States under the leadership of SRON Netherlands Institute for Space Research Groningen. This program is made possible thanks to time guaranteed by a hardware contribution funded by Swiss PRODEX (grant 13911/99/NL/SFe). The work on star formation at ETH Zurich is partially supported by the Swiss National Science Foundation (grants 20-113556 and 200020-121676).
\end{acknowledgements}


\end{document}